\documentclass[12pt,preprint]{aastex}

\usepackage{epstopdf}
\begin{document}

\shortauthors{Anjali et al.}
\shorttitle{Polar Faculae and Polar Magnetic Patches}

\title{The Association of Polar Faculae with Polar Magnetic Patches Examined with {\em Hinode} Observations}

\author{Anjali John K.\altaffilmark{1,2},
Y. Suematsu\altaffilmark{2},
M. Kubo\altaffilmark{2},
D. Shiota\altaffilmark{3} and
S. Tsuneta\altaffilmark{4}
}
\altaffiltext{1}{Graduate University for Advanced Studies, Mitaka, Tokyo 181-8588, Japan; anjali.johnk@nao.ac.jp.}
\altaffiltext{2}{National Astronomical Observatory of Japan (NAOJ), Mitaka, Tokyo 181-8588, Japan.}
\altaffiltext{3}{Solar-Terrestrial Environment Laboratory, Nagoya University, Nagoya 464-8601, Japan.}
\altaffiltext{4}{Institute of Space and Astronautical Science (ISAS), Japan Aerospace Exploration Agency (JAXA), Sagamihara, Kanagawa 252-5210, Japan.}

\begin{abstract}

The magnetic properties of the Sun’s polar faculae are investigated with spectropolarimetric observations of the north polar region obtained by the {\em Hinode} satellite in 2007 September. Polar faculae are embedded in nearly all magnetic patches with fluxes greater than $10^{18}$ Mx, while magnetic patches without polar faculae dominate in the flux range below $10^{18}$ Mx. The faculae are considerably smaller than their parent patches, and single magnetic patches contain single or multiple faculae. The faculae in general have higher intrinsic magnetic field strengths than the surrounding regions within their parent patches. Less than 20\% of the total magnetic flux contributed by the large (${\geq}10^{18}$ Mx) concentrations, which are known to be modulated by the solar cycle, is accounted for by the associated polar faculae.
\end{abstract}


\section{INTRODUCTION}

Polar faculae are bright, small-scale structures visible in white light and in chromospheric lines. They populate higher heliographic latitudes, above 60$^{\circ}$ \citep[e.g.,][]{ok,bl} or 70$^{\circ}$ \citep{an}. \cite{ok} found that the contrast of polar faculae decreases monotonically toward the extreme limb, whereas \cite{bl} reported that the contrast remains constant or may even increase toward the limb.

The number of polar faculae is considered to be a good proxy for the polar magnetic flux. It follows an 11 yr cycle that is anticorrelated with the sunspot cycle \citep{mk,sh,an}. The faculae become visible after the reversal of the polar magnetic field around sunspot maximum, and their numbers reach a maximum at sunspot minimum \citep{shiv}. \cite{ho} and \cite{ok} found that polar faculae have magnetic field strengths in the kilogauss range and that they are unipolar, with the same polarity as the observed global polar field. However, \cite{bl} reported the existence of faculae having polarities opposite to that of the polar field.

The study of the magnetic field in the Sun’s polar regions is important to understand the mechanism of the solar cycle and the origin of the fast solar wind, and the latter’s extension to the interplanetary magnetic field. Information from the polar field is also used to predict the strength of the next solar cycle \citep{sc,ch}. Recent high-resolution observations with the spectropolarimeter \citep[SP;][]{lit} of the Solar Optical Telescope \citep{ichi,shi,sue,tsb} aboard {\em Hinode} \citep{ks} have shown that the polar region has isolated patches of concentrated magnetic field with strengths exceeding 1~kG \citep{tsa,ito}. Magnetic flux concentrations in the polar region separate into two categories: the small-flux concentrations (${<}10^{18}$ Mx) are of mixed polarity with balanced fluxes, whereas the large-flux patches (${\geq}10^{18}$ Mx) scattered over the polar region have, almost without exception, the same polarity \citep{sd}.

\cite{tsa} reported that the kilogauss patches coincide in position with polar faculae, but the true nature of the relationship between polar faculae and the polar magnetic patches remains elusive. In this paper, we investigate how closely the faculae are associated with the magnetic patches by studying their polarity and intrinsic field strength, the inclination of the magnetic field vector with respect to the local normal (zenith angle), and their contribution to the magnetic flux of the observed polar region. We also attempt to determine parameters that control the brightening of polar faculae. These observations correspond to a phase of solar minimum and, hence, to a maximum of polar faculae. At the time of the observations, the polarity of the north polar cap was negative. We hereafter refer to magnetic patches with negative polarity as {\em majority} (dominant) polarity patches and those with positive polarity as {\em minority} polarity patches.

\section{DATA ANALYSIS}
We analyzed the north polar region data taken by the {\em Hinode} SP in 2007 September (Table 1). Each map of the polar region is acquired through several hours of slit-scan observations. The SP recorded the Stokes $I$, $Q$, $U$, and $V$ profiles of the \ion{Fe}{1} 630.15 and 630.25 nm spectral lines, including the nearby continuum. The spectral sampling was 21.55 m\AA. The Stokes spectra were recorded with an integration time of 9.6 s at a spatial resolution of 0$\farcs$32. The pixel size along the slit was 0$\farcs$16.

A least-squares fit using the MILOS code \citep[{\em Mil\/}ne-Eddington Inversion of P{\em o\/}larized {\em S\/}pectra;][]{do} was applied to the Stokes profiles to infer 10 free parameters: three components describing the vector magnetic field (strength, inclination angle, and azimuth angle), the line-of-sight velocity, two parameters describing the linear source function, the ratio of line to continuum absorption coefficients, the Doppler width, the damping parameter, and the stray-light factor $\alpha$. One of the assumptions employed by MILOS is a two-component atmosphere (magnetic and nonmagnetic) in resolved elements, and we are able to obtain the magnetic filling factor, given by $f = 1 - \alpha$.

The 180$^{\circ}$ ambiguity in the transverse magnetic fields was resolved by employing the method of \cite{ito}: The two solutions of the vector magnetic field for each pixel that arise from the ambiguity are obtained in the local coordinates. This provides two different zenith angles, corresponding to the two solutions. We assume that the magnetic field vector is either vertical or horizontal to the local solar surface. The direction is deemed to be vertical if the zenith angle is between 0$^{\circ}$ and 40$^{\circ}$ or between 140$^{\circ}$ and 180$^{\circ}$.

We consider only the vertical magnetic field, because the magnetic field vectors associated with polar patches are nearly vertical to the local surface \citep[e.g.,][]{tsa}. The vertical magnetic flux is defined as $\sum_jB_j f_j \cos i_j A_j$, where $B_j$, $f_j$, $i_j$, and $A_j$ are the intrinsic field strength, magnetic filling factor, zenith angle, and pixel area, respectively, for the $j$th CCD pixel. Since the magnetic field in the polar region is distributed as isolated patches, neighboring pixels with the same polarity are grouped into patches following the method described by \cite{sd}, and each is given a signed identification number based on its polarity.

\subsection{Identification of Polar Facular Pixels within Magnetic Patches}
Magnetic patches at heliographic latitudes above 70$^{\circ}$ with at least 4 pixels, corresponding to the spatial resolution of the telescope, were selected for analysis. Patches with $\mu$ (cosine of the heliocentric angle) less than 0.126, corresponding to a heliocentric angle of 82$\fdg$8, were ignored, since the granular structure is not recoverable beyond this angle.

We picked out magnetic patches and then identified facular pixels inside each patch in the normalized continuum intensity maps. The continuum intensity $(I_C)$ map is obtained by integrating the Stokes $I$ profile outside the absorption lines. The integration ranges were 6300.98--6301.08 and 6302.90--6302.99 \AA. $\langle I_C \rangle$ is a smooth function of the center-to-limb variation (CLV) of the continuum intensity, obtained as follows: A least-squares surface fit using a 5th-order polynomial in $\mu$ \citep[following][]{pi} is performed on the $I_C$ map. The fitted map is then subtracted from the original, and the standard deviation $\sigma_{0}$ of the difference is calculated. We then remove the bright and dark features from the original image using a $\pm3\sigma_{0}$ cutoff, and a fit with same functional form is performed to obtain a CLV function unaffected by the presence of faculae. The normalized intensity is defined as ${I_C}/{\langle I_C \rangle}$, where $I_C$ and $\langle I_C \rangle$ are the continuum intensity and the intensity averaged over the same $\mu$-value, respectively.

Figure 1 shows the normalized intensity as a function of $\mu$. The spikes in the plot are attributed to polar faculae, while the smaller fluctuations about 1.0 are ascribed to granulation. Within each magnetic patch, pixels having intensity greater than or equal to a given threshold are classified as belonging to polar faculae. The threshold to identify facular pixels varies with $\mu$. The standard deviation $\sigma$ of the normalized intensity (with respect to the smooth CLV function) is derived at each $\mu$ with a window of size 0.01, and the $\sigma$'s are fitted with a 3rd-order polynomial in $\mu$ for each of the six observations. The threshold to detect faculae is set to 4 $\sigma$, which is well above the intensity fluctuations due to granulation, as shown in Figure 1 (thick solid line).

\section{Results}

We find that polar faculae, as defined above, are present in most (55\%) of the majority polarity magnetic patches with flux greater than $10^{18}$ Mx (Figure 2). Concentrations of large magnetic flux, which vary with the solar cycle \citep{sd}, harbor polar faculae. Numerous patches without faculae also exist in the polar region; at fluxes below $10^{18}$ Mx, patches without faculae far exceed those with faculae.

\subsection{Majority Polarity Patches}
Here we focus on majority polarity magnetic patches with polar faculae. We find that the patches with faculae are not uniformly bright but instead contain smaller faculae. Figure 3 shows two examples of patches with polar faculae. In the first (panel (a)), four faculae are identified in one patch. They differ in size, intensity, and magnetic flux. In most cases, the facular islands enclose the local flux maximum within the magnetic patch (e.g., the two uppermost faculae in panel (b)). In the second patch (panel (c)), only one facula is identified. These examples reflect the general case that the number of polar faculae associated with each patch ranges from one to a few. The faculae have fine structure, with a core and an extended halo region (represented by arrows A and B, respectively, in Figure 3).

Each magnetic patch is taken to comprise two components: a polar facular region and a nonfacular region. As polar faculae are small structures compared with the nonfacular region within a patch, it is reasonable to use the peak value of ${I_C}/{\langle I_C \rangle}$ rather than its average value to characterize the faculae, to better capture localized behavior. We refrain from plotting the peak values against flux for nonfacular regions, as it is possible that our threshold might misclassify some true facular pixels as nonfacular. Such errors, if any, could be minimized by taking the average of the normalized continuum intensity. The top panel of Figure 4 is a scatter plot of the peak intensity inside the polar faculae versus their magnetic flux. For each patch, the peak intensity was obtained by considering all the facular pixels in the patch, and the magnetic flux was calculated by summing the flux of all the facular pixels within the patch. There is a clear positive correlation between the peak intensity and the magnetic flux. For magnetic fluxes ${\leq}10^{17}$ Mx, the peak intensity lies between 10\% and 20\% above the average, beyond which it increases from 20\% to 44\%. This trend can be explained with the results from \cite{venk}, who discovered that convective instability is less efficient for flux tubes with lower magnetic flux (${\leq}10^{17}$ Mx). This is a consequence of efficient thermal coupling between smaller flux tubes and the ambient nonmagnetic atmosphere, which hinders convective collapse. Observational confirmation of this finding was given by \cite{sol} using quiet-Sun data near disk center. Hence, faculae need to have minimal magnetic flux to be visible as distinct magnetic features.

In the bottom panel of Figure 4, we map the normalized intensity averaged over facular (crosses) and nonfacular (triangles) regions against their respective fluxes. The average intensity of the faculae increases with magnetic flux, as expected. A comparison of the two panels makes clear that for faculae, peak intensity is better correlated than average intensity with the magnetic flux. The bottom panel also shows that the average intensity of nonfacular regions decreases with magnetic flux. This bimodal distribution is essentially the same for different $\mu$-ranges. For a given flux value, there exist both bright facular and darker nonfacular regions. Hence, magnetic flux cannot be used to discriminate between facular and nonfacular regions within a magnetic patch. Note that the branching becomes evident at fluxes greater than $10^{17}$ Mx. The magnetic patch with large magnetic flux is larger in size and hence it is likely that the nonfacular region in such magnetic patch will mostly be occupied by granules . We think this is the reason why the average intensity of the nonfacular region approaches unity with increasing magnetic flux.

In Figure 5, we plot the probability distribution functions (PDFs) of intrinsic field strength averaged over facular and nonfacular pixels within each magnetic patch. The average field strength of the polar faculae displays a broad peak at 700--1000 G. The nonfacular regions exhibit a peak around 400 G. Since the degree of the depression of the visible surface in a flux tube depends on the intrinsic magnetic field strength, this sharp difference is consistent with the picture set forth by \cite{spr} (see Section 4).

Figure 5 also shows, however, that some faculae have very small average intrinsic magnetic field strengths. The field strength may decrease toward the limb as a result of our viewing greater heights in the atmosphere, thus leading to weaker observed fields. In this case, the low values correspond to those from smaller $\mu$. Our examination did not identify any trend in this relationship. We also found that the average intensity of these faculae does not show any dependence on the intrinsic field strength averaged over the respective facular pixels. It is not clear whether these represent different evolutionary phases of faculae, and so to clarify the situation, it will be necessary to examine the temporal evolution of the magnetic patches.

The PDFs of the zenith angles of the magnetic field vectors averaged over facular and nonfacular pixels within each magnetic patch are shown in Figure 6. The vectors of both regions are nearly vertical to the local normal, but those of nonfacular regions appear to be slightly inclined. The PDF of the zenith angle has a peak at approximately 169$^\circ$ for facular regions and a maximum around 165$^\circ$ for nonfacular regions. This small but clear difference may reflect a tendency of faculae to be located inside the canopy-like structure \citep{tsa} of magnetic patches.

The PDF of the average filling factor for polar faculae shows a broad distribution ranging from 0.1 to 0.6 (Figure 7), while the nonfacular regions have a distinct peak at $\sim$0.2. Although some polar faculae have a very high filling factor, there exists no significant difference between facular and nonfacular regions within magnetic patches as compared with the intrinsic field strength or the zenith angle of the magnetic field vector, as is evident from Figures 5 and 6. Figure 8 depicts the variation with $\mu$ of the average intensity of polar facular regions within each magnetic patch. The intensity decreases gradually toward the limb, consistent with \cite{ok}.

\subsection{Minority Polarity Patches}
We find that polar faculae are associated with minority polarity patches as well, which is to say, they are polarity independent. However, faculae with minority polarity are very limited in number, and most have magnetic fluxes below $10^{18}$ Mx. Figure 9 is a scatter plot of the peak intensity of minority polarity faculae as a function of their magnetic flux. This should be compared with Figure 4 (top panel), for the majority patches. Peak intensity appears to have a positive correlation with magnetic flux. The number of minority patches detected (82) is too small to allow us to make a definite conclusion.

\section{Summary and Discussion}

We have found that polar magnetic patches have substructure, with one or more small faculae embedded in the much larger patches. The faculae appear to be a subset of magnetic patches. Their shapes inside the patches are irregular. Most of the large magnetic concentrations, which have cyclic behavior, host faculae. We also found that faculae exhibit a tendency to have higher intrinsic magnetic field strengths compared with the nonfacular regions inside the associated magnetic patches. Table 1 lists the ratio of the magnetic flux of the faculae to that of the patches (including those without polar faculae). We find that less than 20\% of the total magnetic flux from the large concentrations is accounted for by the associated faculae. It is important to study the cyclic variation of the facular flux contribution to the large concentrations in order to understand the exact relationship between polar faculae and the solar cycle.

We found that minority polarity faculae also exist in the polar region. The number of minority polarity faculae might depend on the strength of the unipolar field in the polar region. Hence, for weak solar cycles, care should be exercised in assuming that the count of polar faculae (which have been presumed to be unipolar in most previous studies) is linearly related to the total signed polar magnetic flux.

We expected in this investigation to find controlling parameters and/or environment that switch polar faculae on or off. Intrinsic magnetic field strength and magnetic flux indeed correlate well with the existence of polar faculae, as shown in Section 3.1, but the correlation is somewhat ambiguous, as shown respectively in Figures 5 and 2. Polar faculae possess stronger and more vertical fields than their surroundings within a magnetic patch. This tendency may be due to the polar faculae being located near the patch centers, which have a canopy-like magnetic structure \citep{tsa}. Thus, the observation might simply reflect the magnetic structure.

Our observation that faculae possess strong vertical magnetic fields with average intensity decreasing toward the limb is consistent with the hot-wall model \citep{spr}, which attributes the enhanced brightness of faculae to a depression in the visible surface caused by magnetic pressure, allowing an enhanced view of the hot wall of the flux tube at oblique angles.

We do not have information on the evolution of polar faculae during the development of the parent magnetic patches from just the snapshot slit-scan observations. It remains necessary to investigate the temporal evolution of magnetic patches and polar faculae to further constrain the properties of faculae.

\acknowledgements

We are grateful to the anonymous referee whose supporting recommendations helped much in improving the manuscript. We also thank Robert Cameron for valuable discussions and comments. A. J. K. thanks the Ministry of Education, Culture, Sports, Science, and Technology (MEXT) of Japan for financial support through its doctoral fellowship program for foreign students and the Sokendai for an associate researcher's grant. \emph{Hinode} is a Japanese mission developed and launched by ISAS/JAXA, with NAOJ as domestic partner and NASA and STFC (UK) as international partners. It is operated by these agencies in cooperation with ESA and NSC (Norway). Use of NASA's Astrophysics Data System is gratefully acknowledged.


\bibliography{bib}

\begin{figure}
\epsscale{0.50}
\plotone{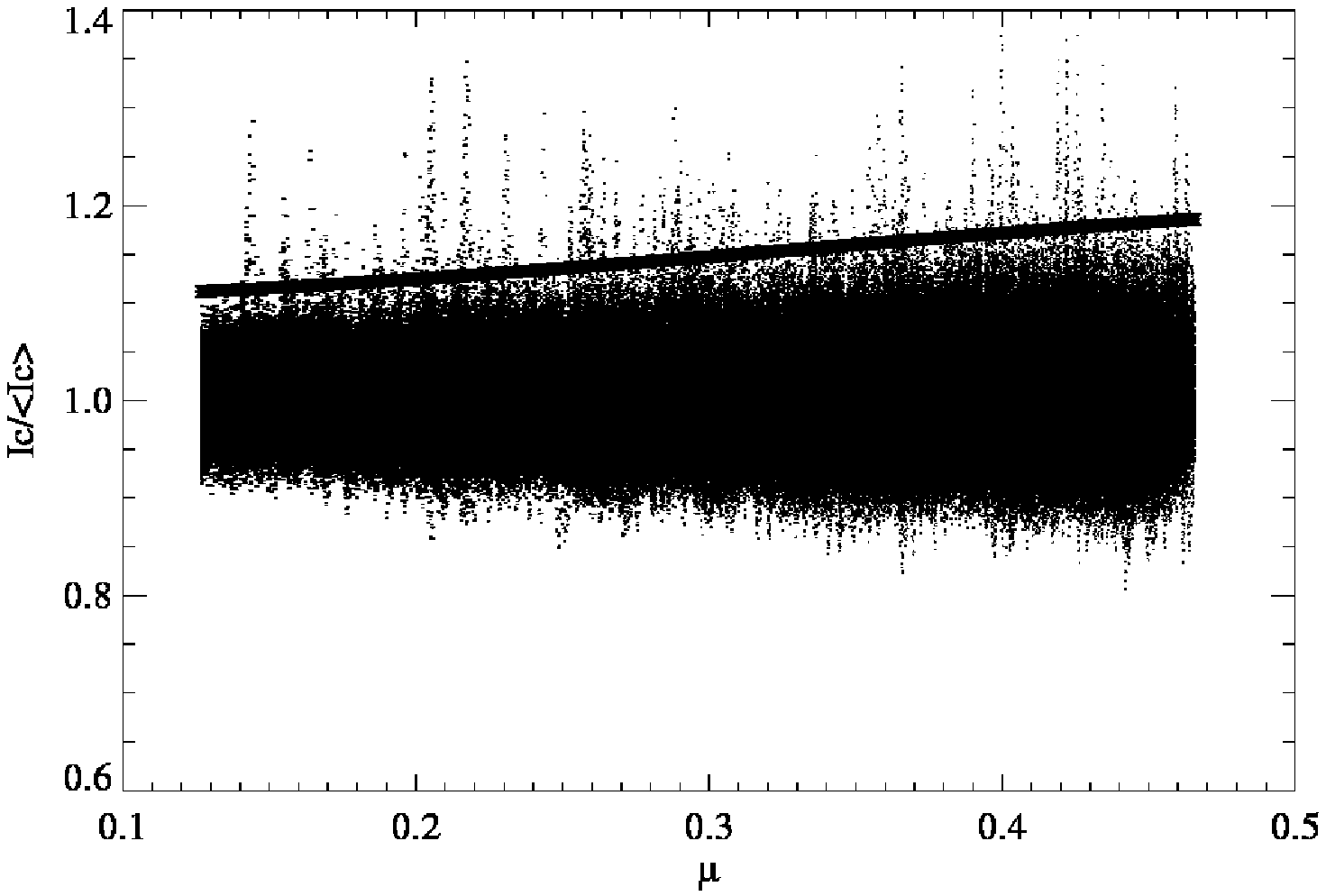}
\caption{Normalized continuum intensity vs. $\mu$ (cosine of the heliocentric angle) for all pixels (2007 September 4). The $\mu$-values increase with distance from the limb. In this plot, the polar faculae appear as intensity spikes. Both granular and facular intensities decrease toward the limb. The thick solid line indicates the 4 $\sigma$ fluctuation threshold used to pick out faculae. The value 1.0 is the quiet-Sun intensity averaged over same $\mu$-value (see Section 2.1 for details).}
\end{figure}

\begin{figure}
\epsscale{0.50}
\plotone{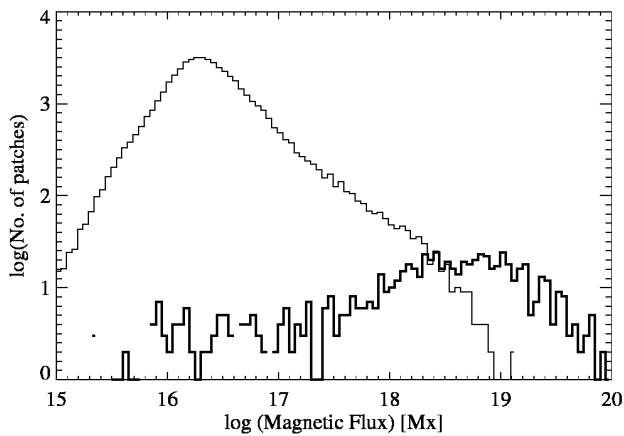}
\caption{Histogram of the magnetic flux of the majority polarity patches. The thick line represents patches that contain one or more faculae, and the thin line shows patches without faculae. The values shown on the $x$-axis correspond to flux per patch. The magnetic flux of a patch was obtained by integrating the flux of all pixels within the patch.}
\end{figure}

\begin{figure}
\epsscale{0.65}
\plotone{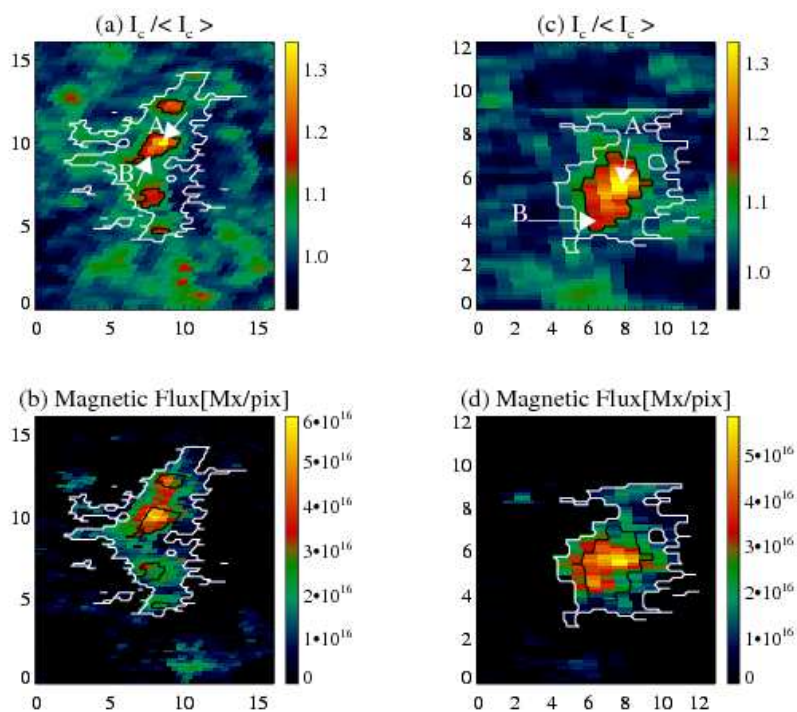} 
\caption{Two examples of magnetic patches associated with polar faculae: (a, c) normalized continuum intensity; (b, d) magnetic flux in units of maxwells per pixel. White contours enclose the patches, and black contours enclose the polar faculae within the patch. The $x$- and $y$axes are in arcseconds. For patch 1 (left), the magnetic flux is $3.07\times10^{19}$ Mx, and $\mu\approx0.26$; for patch 2 (right), the flux is $2.24\times10^{19}$ Mx, and $\mu\approx0.21$. Arrows indicate core (A) and extended (B) regions of the faculae.}
\end{figure}

\begin{figure}
\epsscale{1.05}
\plottwo{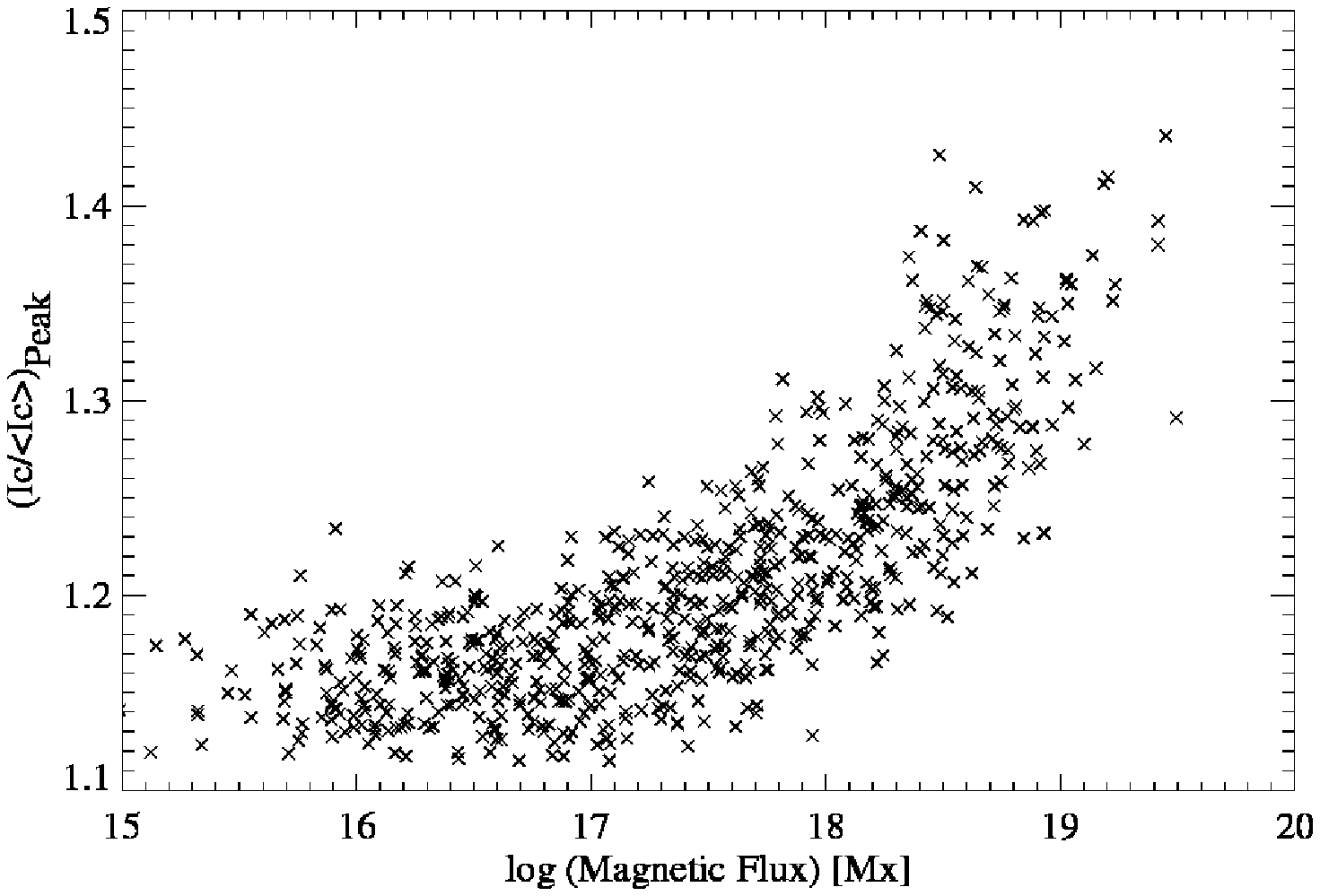} {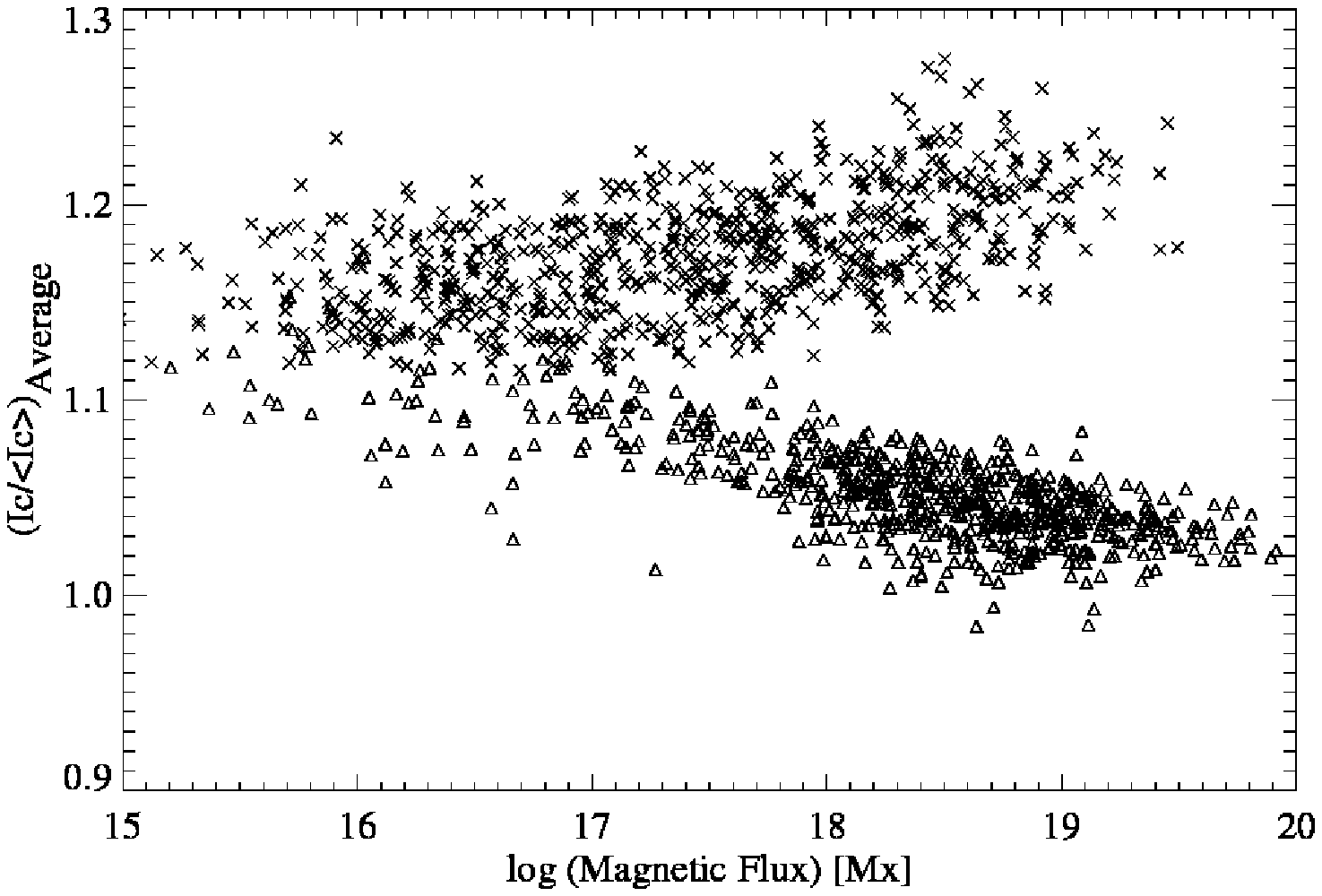}
\caption{Left: Peak value of the normalized continuum intensity ${I_C}/{\langle I_C \rangle}$ of majority polarity polar faculae vs. magnetic flux summed over facular pixels within each magnetic patch. Right: Scatter plot of the average value of ${I_C}/{\langle I_C \rangle}$ of majority facular (crosses) and nonfacular (triangles) regions as a function of magnetic flux integrated over the respective pixels within each patch. Peak and average intensities of polar faculae are calculated over all the facular pixels identified within each patch. For each patch, the average intensity of the nonfacular region is determined by taking the mean of intensity of all the pixels outside faculae within that patch.}
\end{figure}

\begin{figure}
\epsscale{0.50}
\plotone{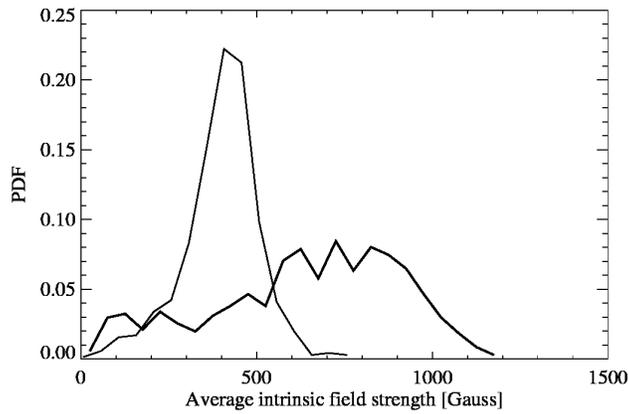}
\caption{PDFs of average intrinsic field strength of polar facular (thick solid line) and nonfacular (thin solid line) regions within magnetic patches. The bin size is 50 G.}
\end{figure}

\begin{figure}
\epsscale{0.50}
\plotone{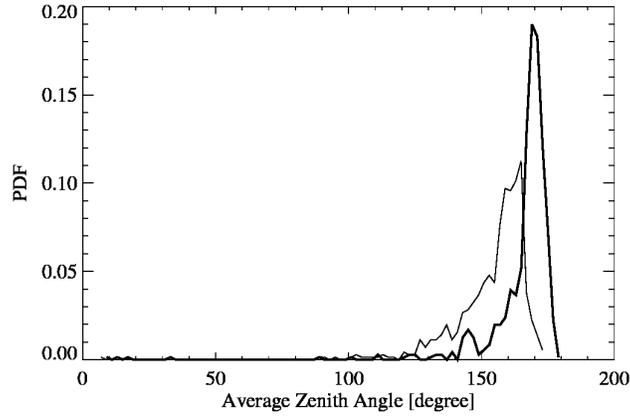}
\caption{Same as Figure 5, but for average zenith angle. The bin size is 2$^\circ$.}
\end{figure}

\begin{figure}
\epsscale{0.50}
\plotone{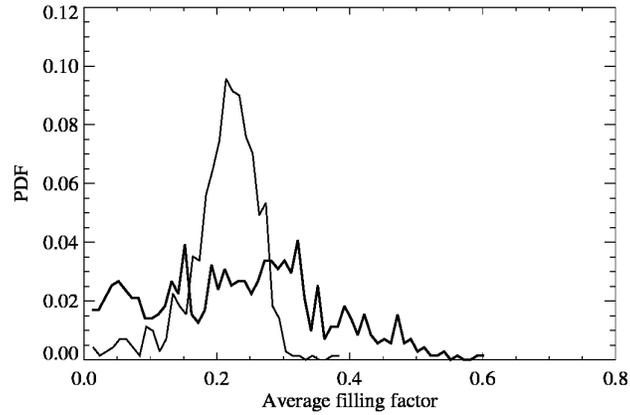}
\caption{Same as Figure 5, but for average filling factor (see Section 2) The bin size is 0.01.}
\end{figure}

\begin{figure}
\epsscale{0.50}
\plotone{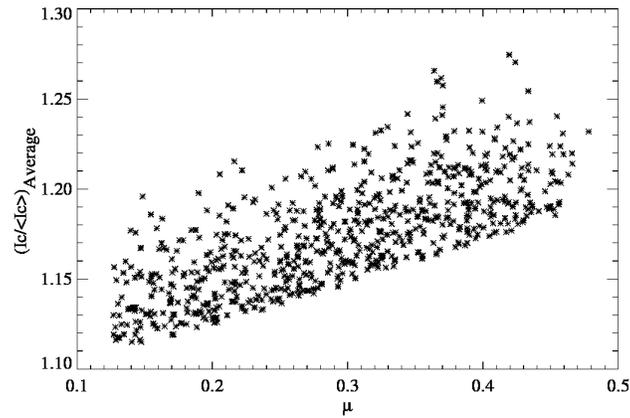} 
\caption{Scatter plot of the average normalized continuum intensity of polar faculae as a function of $\mu$. The lower boundary of the distribution is due to the 4 $\sigma$ threshold.}
\end{figure}

\begin{figure}
\epsscale{0.50}
\plotone{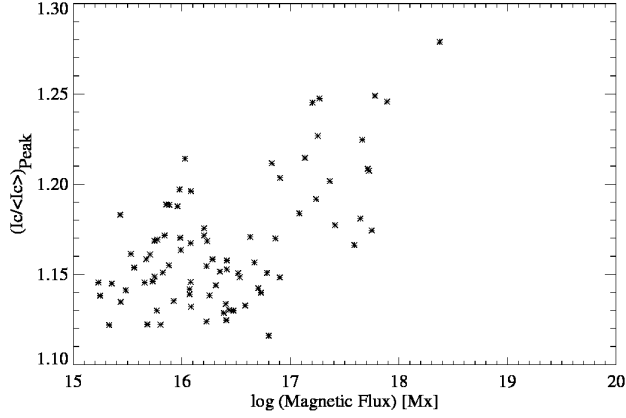} 
\caption{Same as the top panel of Figure 4, but for minority polarity faculae.}
\end{figure}

\begin{deluxetable}{cccccccccc}
\tabletypesize{\scriptsize}
\tablecolumns{7}
\tablewidth{0pc}
\tablecaption{{\em Hinode} Data Set}
\tablehead{
\colhead{Date} &\colhead{Time} &\colhead{Exp. time} &\multicolumn{4}{c}{Field of View\tablenotemark{a}} &\colhead{TMF\tablenotemark{b}} &\colhead{PFF/TMF\tablenotemark{c}} &\colhead{MPF\tablenotemark{d}}\\
\colhead{} &\colhead{(UT)} &\colhead{(S)} &\colhead{Xmin} &\colhead{Xmax} &\colhead{Ymin} &\colhead{Ymax} &\colhead{$\times 10^{20}$ Mx} &\colhead{(\%)} &\colhead{$\times 10^{19}$ Mx}\\
\colhead{} &\colhead{} &\colhead{} &\colhead{(arcsec)} &\colhead{(arcsec)} &\colhead{(arcsec)} &\colhead{(arcsec)} &\colhead{} &\colhead{} &\colhead{}}
\startdata
2007 Sep 04 &18:13-23:42 &9.6 &-194.86 &125.12 &841.81 &1005.65 &9.06 &16.74&3.01\\ 
2007 Sep 06 &00:28-05:57 &9.6 &-197.28 &122.72 &842.43 &1006.27 &12.4 &18.2&3.71\\ 
2007 Sep 08 &01:07-06:36 &9.6 &-186.08 &133.92 &840.67 &1004.51 &13.15 &15.2&2.67\\ 
2007 Sep 09 &01:05-06:35 &9.6 &-186.72 &133.28 &842.35 &1006.19 &10.0 &11.26&2.68\\ 
2007 Sep 10 &01:15-06:39 &9.6 &-200.00 &115.00 &841.47 &1005.31 &12.3 &16.5&2.10\\
2007 Sep 16 &01:23-06:52 &9.6 &-183.52 &136.42 &832.25 &996.09 &10.56 &15.0&3.75
\enddata
\tablenotetext{a}{In heliographic coordinates. $X$ is directed toward solar west and $Y$ toward solar north. The origin of the coordinate system is Sun center.}
\tablenotetext{b}{Total vertical magnetic flux contributed by all the majority polarity patches (with flux ${\geq}10^{18}$ Mx), both with and without polar faculae.}
\tablenotetext{c}{PFF denotes the total magnetic flux from polar faculae associated with majority polarity patches with flux ${\geq}10^{18}$ Mx detected in this study.}
\tablenotetext{d}{Total minority polarity flux contributed by all the minority polarity patches (with flux ${\geq}10^{18}$ Mx), both with and without polar faculae.}
\end{deluxetable}
\end{document}